\documentclass[twocolumn]{aa}      
\usepackage{graphicx}      

\begin{document}      
   \title{Modelling Chromospheric Line Profiles in NGC2808: Evidence of 
   Mass Loss from RGB Stars 
   \thanks{Based on observations collected at the European Southern      
   Observatory, Chile, during FLAMES Science Verification}}      
      
      
   \author{P.J.D. Mauas\inst{1},
         C. Cacciari\inst{2},     
	 and     
	 L. Pasquini\inst{3}     
	 }      
      
   \offprints{C. Cacciari}      
      
   \institute{ 
   Instituto de Astronom\'\i a y F\'\i sica del Espacio, Buenos Aires, Argentina\\
   \email{pablo@iafe.uba.ar}
   \and
   INAF-Osservatorio Astronomico di Bologna, via Ranzani 1, 40127 Bologna, 
   Italy \\      
   \email{carla.cacciari@bo.astro.it}   
   \and     
   European Southern Observatory, Karl-Schwarzschild-Str. 2, D-85749 
   Garching b. M\"unchen, Germany \\     
   \email{lpasquin@eso.org}	     	        
}       
     
   \date{}      
      
   \abstract{In this study we test the possibility that the asymmetry
in the profiles of the H$\alpha$ and Ca {\sc ii} K lines in red giant stars 
is due to the presence of an active chromosphere rather than to mass loss.  
To this end, we compare line profiles computed using relevant model 
chromospheres to profiles of the H$\alpha$ and Ca {\sc ii} K lines observed 
in five red giant stars of the globular cluster NGC 2808. 
The spectra were taken with FLAMES during Science Verification, using the 
UVES mode at high resolution (R=43,000) for the H$\alpha$ line, and GIRAFFE 
in MEDUSA mode (R=20,000) for the Ca {\sc ii} K line. 
We find that the observed profiles are better described if a negative 
(outward) velocity field is included in the model chromospheres. 
This leads to mass loss rates of a few 10$^{-9}$ M$_{\odot}$ yr$^{-1}$, 
very close to the requirements of the stellar evolution theory.

}

 \authorrunning{Mauas et al.}      
 \titlerunning{Chromospheres and mass loss in NGC 2808 red giant stars}      
 \maketitle      
%

\section{Introduction}      

It has long been known that 
the stellar evolution theory requires that some mass 
($\sim$ 0.1-0.2 M$_{\odot}$) 
be lost by Population II stars during the evolutionary phases 
preceding the horizontal branch (HB) phase, in order to account for 
(1) the observed morphologies of the HB in globular clusters (GC), and  
the maximum luminosity reached by stars on the asymptotic giant branch 
(Castellani \& Renzini 1968; Iben \& Rood 1970; Rood 1973; 
Martinez Roger \& Paez 1987; Jorgensen \& Thejll 1993; Ferraro et al. 1998; 
Catelan 2000; Soker et al. 2001), 
and (2) the pulsational properties of the RR Lyrae variables (Christy 1966; 
Fusi Pecci et al. 1993; D'Cruz et al. 1996). 

Theoretical estimates of mass loss rates at the tip of the red giant branch 
(RGB) are a few times $10^{-8}$ M$_{\odot}$ yr$^{-1}$ (Fusi Pecci \& 
Renzini 1975, 1976; Renzini 1977). More recently, Mullan \& MacDonald (2003) 
find that the evolution through the ``bump'' in the luminosity function of 
RGB stars might be associated with the onset of a cool 
massive wind. In a study of RGB stars in the halo of the Milky Way, 
de Boer (2004) estimates that on average $\sim$ 0.3 M$_{\odot}$ per star 
are lost by metal-poor red giants, and would contribute a sizeable fraction 
of the observed H {\sc i} gas falling in from the halo.  
  
So, a few tens of solar masses of intracluster matter are expected 
to be produced in any given GC. This matter should accumulate in the 
central regions, especially of the more massive clusters, in absence of 
sweeping mechanisms between Galactic plane crossings.      
However, efforts to obtain evidence of intracluster matter  
have been largely unsuccessful: diffuse gas in GC's  was detected 
only as an upper limit and well below 1 $M_{\odot}$ (Roberts 1988; 
Smith et al. 1990; Faulkner \& Smith 1991; Freire et al. 2001).

\begin{table*}      
\begin{center}            
\caption{Physical and atmospheric parameters for the RGB stars in NGC2808
used for the present study. The temperatures, luminosities and gravities 
listed in columns 3-5 are derived from the (V--K) colours (see Sect. 2). 
The expansion velocities and mass loss rates in columns 11
and 12 are discussed in Sect. 4.1. 
}
\label{t:list}       
\begin{tabular}{rccccccccccc}      
\hline\hline      
\\      
Star ID  &M$_V$&T$_{\rm eff}$(K) & R$_{\ast}$/R$_{\odot}$ & log L$_{\ast}$/L$_{\odot}$ &log g$_{\ast}$ & 
H$\alpha$ shift&B/R & $K_3$ shift & B/R & V$_{exp}$ & $\dot{M}$  \\   
         &     &                 &      &      &     & km~s$^{-1}$ & H$\alpha$ & 
km~s$^{-1}$ & $K_2$ & km~s$^{-1}$ & $M_{\odot} yr^{-1}$ \\   
\\          
\hline      
\\ 
37872 & -1.940 & 4015 & 67 & 3.028 & 0.71 & -2.57 & $<$1 & -5.03 & $<$1 & 15 & 1.1 10$^{-9}$ \\
47606 & -2.154 & 3839 & 92 & 3.218 & 0.44 & -0.31 & $>$1 & -6.25 & $>$1:& 15 & 1.1 10$^{-10}$\\
48889 & -2.249 & 3943 & 84 & 3.188 & 0.52 & -3.48 & $<$1 & -5.19 & $<$1 & 53 & 3.8 10$^{-9}$\\
51454 & -2.144 & 3893 & 85 & 3.177 & 0.51 & -3.95 & $>$1 & -7.70 & $<$1 & 10 & 0.7 10$^{-9}$\\
51499 & -2.155 & 3960 & 79 & 3.142 & 0.57 & -4.39 & $<$1 & -6.25 & $<$1 & 18 & 1.2 10$^{-9}$\\
\\     
\hline      
\end{tabular}      
\end{center}      
\end{table*}      

Direct detection of mass loss from individual metal-poor RGB stars, either 
in the field or in GCs, has been attempted with various observational 
approaches. 
Infrared excess, indicative of dusty circumstellar envelopes, has been 
detected with ISOCAM in $\sim$ 15\% of the globular cluster RGB (or AGB) 
stars in the $\sim$ 0.7 mag brightest interval (M$_{\rm bol} \le -2.5$)      
(see Origlia et al. 2002, and references therein).  
Spectroscopic surveys of a few hundred GC red giants      
(Cohen 1976, 1978, 1979, 1980, 1981; Mallia \& Pagel 1978;      
Peterson 1981, 1982; Cacciari \& Freeman 1983; Gratton et al. 1984)      
revealed H$\alpha$ emission wings in a good fraction of stars along the 
uppermost  1.25 mag interval of the RGB. 
This emission, generally asymmetric and possibly variable on a short-time 
scale, was initially interpreted as evidence of an extended atmosphere, 
until Dupree et al. (1984) demonstrated that it could arise naturally in 
a static stellar chromosphere, in analogy with Reimers (1975, 1977, 1981) 
speculations on Population I red giants.   

Profile {\em asymmetry} and {\em coreshifts} of chromospheric lines, such as 
the Na {\sc i} D and Ca {\sc ii} K lines, in addition to H$\alpha$, were 
then considered as possible indicators of mass motions.  
Several red giants in GCs were found to exhibit small negative velocity      
shifts in the cores of these lines (Peterson 1981; Bates et al. 1990, 1993; 
Dupree et al. 1994; Lyons et al. 1996), as did metal-poor field giants which 
might be taken as the field counterparts of GC giants (Smith et al. 1992; 
Dupree \& Smith 1995). 
These coreshifts are all $\le$ 15 km~s$^{-1}$, i.e. much smaller than the 
escape  velocity from the stellar photosphere ($\sim$ 50-60 km~s$^{-1}$). 
  
However, in those cases where the Mg {\sc ii} $\lambda$2800 $h$ and $k$ 
lines were also available for metal-poor field or GC RGB stars  
(Dupree et al. 1990a, 1990b, 1994; Smith \& Dupree 1998),  
the asymmetries in the Mg {\sc ii} lines, under certain 
assumptions,  provided some evidence of  
a stellar wind with a terminal velocity as high as 
$\sim$ 50--150 km~s$^{-1}$,  exceeding both the stellar photospheric 
escape velocity and the escape velocity from many cluster cores 
($\sim$ 20-70 km~s$^{-1}$; Webbink 1981). 
These values of velocity, along with appropriate values of stellar 
radius and chemical abundance, would be consistent with mass loss rates 
of a few 10$^{-11}$ to a few 10$^{-9}$ M$_{\odot}$ yr$^{-1}$
(Dupree et al. 1990b; 1994),  leading to a total mass loss 
of up to $\sim$ 0.2 M$_{\odot}$ over a time of $\sim 2 \times 10^{8}$ yr
on the RGB, in general agreement with the expectations      
of the stellar evolution theory.  

Further observational evidence of mass outflows from individual 
metal-poor red giants was provided by Dupree et al. (1992) and 
Smith, Dupree \& Strader (2004) using the He {\sc i} $\lambda$10830.3 
absorption line on five field metal-poor red giants and six RGB stars 
in the GC M13. This  weak line forms higher in the atmosphere than H$\alpha$ 
or the Ca {\sc ii} and Mg {\sc ii} emission cores and is, therefore,
a better probe of the outer regions of the atmosphere where the wind begins 
to accelerate. The study by Smith et al. (2004) has revealed the 
possible presence of mass outflows in fainter (and hotter) RGB stars 
than previously monitored by chromospheric lines in the UV 
and optical range. 

If we restrict our attention only to studies in the optical range,   
the most recent piece of work was done by Cacciari et al. (2004, 
hereafter C04), who used FLAMES (Pasquini et al. 2002)  
spectra centered on Ca {\sc ii} K, 
Na {\sc i} D and   H$\alpha$ lines of $\sim$100 RGB stars in NGC 2808. 
Their purpose was to monitor the profiles of these lines along the RGB 
in search of asymmetries and core shifts, taken to indicate mass motions 
in the atmospheres.
Indeed, a good fraction of the brightest RGB stars was found to show 
these features, with negative (outward) velocities of $\le$ 10 km~s$^{-1}$. 
However, these velocities are quite low and might be accounted for by the 
presence of local hydrodynamic effects in the chromosphere  
(Dupree et al. 1984; Smith \& Dupree 1988; Dupree \& Smith 1995).    
One could fairly summarize the current status by stating that the 
long-sought direct evidence for mass loss in Population II RGB stars is 
still quite elusive.

The purpose of the present study is to compare the observed profiles of 
the Ca {\sc ii} K and H$\alpha$ lines in a few RGB stars in the
GC NGC 2808 with theoretical profiles from suitable model chromospheres, and 
verify whether a static chromosphere can account for the observed features or 
a velocity field needs to be considered. 
In the latter case, we estimate quantitatively if the adopted field is 
consistent with the amount of mass loss expected from theory.
 
In Sect. 2 and 3 we describe the data and the models we have used, in Sect. 4 
the results of our analysis are discussed, and Sect. 5 contains a summary of 
this work and our conclusions.

\section{The Data} 

We list in Table \ref{t:list} the five RGB stars of NGC2808 we have selected 
from the C04 sample for the present analysis. They have been 
selected among the brightest and coolest stars with clear evidence of 
emission in the H$\alpha$ wings and at the bottom of the Ca {\sc ii} K 
absorption core (K$_2$ reversal). 
The temperatures and luminosities listed in Table \ref{t:list} are derived 
from the (V--K) colours, and 
the gravities assume a mass of 0.85 M$_{\odot}$ for these stars (see C04 and 
Carretta, Bragaglia \& Cacciari 2004 for more details). 
Columns 6 and 8  show the shifts 
with respect to photospheric lines of the H$\alpha$ and Ca {\sc ii} K$_3$  
absorption line cores, respectively. Columns 7 and 9 show the asymmetry of 
the H$\alpha$ and K$_2$ emission using the ratio B/R 
of the respective blue and red wings.
As found in all previous studies, the emission wings show the characteristic 
signature of downflow (i.e. B/R$>$1) about as often as that of outflow 
(B/R$<$1), suggesting that the emission could be produced in
a locally non-static chromosphere.     
The consistently negative shifts of the absorption cores are in 
principle better indicators of a net outflow, however the low velocity of 
these features might also be compatible with normal chromospheric activity.

\section{The Model Chromospheres}  

\begin{figure}      
\begin{center}     
\includegraphics[width=9cm]{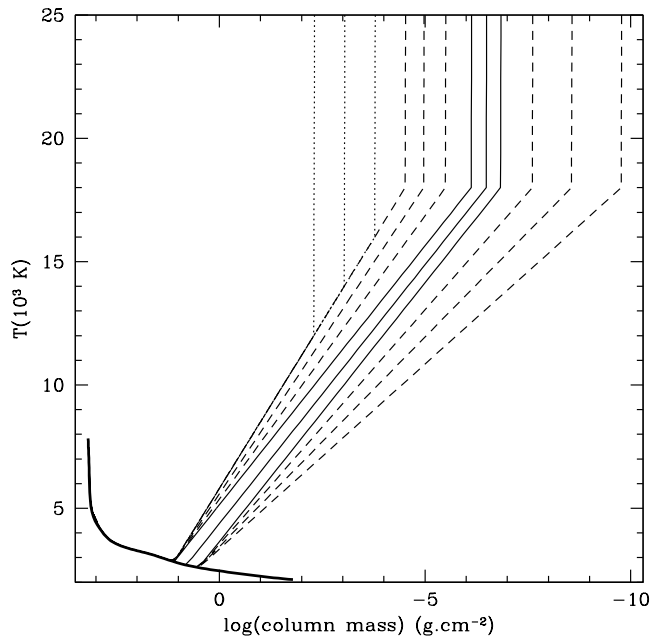}       
\caption{Examples of our chromospheric models. The thick
  line is Kurucz' model for log(g) = 0.5 and $T_{eff}=4000$ K.
  Different models are shown with different line types for clarity.}
\label{f:exa}      
\end{center}     
\end{figure}

For the photospheric temperature distribution we used the
model computed by Kurucz (2005) for log(g) = 0.5
and $T_{eff}=4000$ K. 
Although there are differences in these parameters for the
stars in our sample, this model is the closest one in all cases 
(except perhaps for star 47606, for which the model with $T_{eff}=3750$ 
K is  marginally closer).
On this model, we overimposed a
chromospheric rise, with $T$ increasing linearly with negative  
log(m). In this way, our chromospheric models depend on
three free parameters: i) the column mass log(m) where the chromosphere
starts, {\it i.e.} where the temperature distribution departs
from Kurucz' model and where, therefore, the temperature has
its minimum value; ii) the slope of the chromospheric $T$ vs
log(m) distribution; and iii) the maximum temperature to which
the chromospheric rise extends, which relates directly with
the value of log(m) where the chromosphere ends. Upon this 
chromosphere, we added a ``transition region'' where the
temperature rises abruptly up to $1.5 \times 10^5$ K, to
assure convergence of the calculations. The structure of this 
``transition region'' has no influence in the computed
profiles. As a matter of fact, the computed profiles are 
unaffected by the structure of the regions where the 
temperature is higher than $1. \times 10^4$ K, as 
we see below. Examples of the different models we have computed are shown in
Fig. \ref{f:exa}.

For every $T$ vs log(m) distribution, we solved the
non-LTE radiative transfer and the statistical and
hydrostatic equilibrium equations, using the program Pandora 
(Avrett \& Loeser 1984). 
Assuming a gravity log(g) = 0.5,  
a turbulent velocity varying from 2 to 20 km/s and
a metallicity[Fe/H]=--1.14
all elements scaled solar, we self-consistently 
computed non-LTE populations for 10 levels of H (for details
on the atomic models, see Falchi \& Mauas 2002),  
13 of He {\sc i} (see Mauas et al. 2004),  
9 of C {\sc i}, 15 of Fe {\sc i}, 8 of Si {\sc i} (see
Cincunegui \& Mauas 2001),  
8 of Ca {\sc i} and Na {\sc i} and  6 of Al {\sc i} (see
Mauas, Fern\'andez Borda \& Luoni 2002), 
and 7 of Mg {\sc i} (see Mauas, Avrett \& Loeser 1988).  
In addition, we computed 6 levels of He {\sc ii} (see Mauas
et al. 2004)  
and Mg {\sc ii}, and 5 of Ca {\sc ii} (see Falchi \& Mauas 2002). 
For every species under
consideration we included all the bound-free transitions and
the most important bound-bound transitions. Ly-$\alpha$, the
Ca {\sc ii} H and K and Mg {\sc ii} $h$ and $k$ lines were all 
computed with a full partial redistribution treatment. 
A test was performed assuming helium abundance Y=0.23 
and $\alpha$-element enhancement [Ca/Fe]=+0.3, that are
more appropriate for a GC stellar population, and no significant
difference was detected in the results discussed in
the following section. 
We also performed several test runs varying the gravity 
within the observed range, and found that the emitted profiles 
were not affected. In our calculations we assumed a plane-parallel 
atmosphere, for simplicity. However, once the final model was obtained, 
we computed the emitted profiles assuming a spherically symmetric
atmosphere, and also in this case we found no significant changes in 
the emitted profiles with respect to the plane-parallel approximation. 
The abundances adopted in the present study are taken from the 
detailed chemical analysis of the same stars performed by Carretta (2006).

\section{Analysis and Results}

The observed profiles for three of our stars (48889, 51454 and 
51499) are very similar, differing mainly in the amount of
asymmetry in the K$_2$ and H$\alpha$ emission wings. 
The profiles for the other two stars (37872 and 47606) are 
also similar and differ mainly in the asymmetry.
Furthermore, the asymmetries are not very large and
can be considered as a second order change in the
profiles. For these reasons, we computed two atmospheric models, one (model
A) to fit the profiles of the first three stars, and the other
(model B) to fit the profiles of the remaining two stars. 

Once the models that gave the best fit to the symmetric profiles were
obtained, we applied to each of them a velocity field, computed the resulting 
profiles, and compared them with the observed asymmetries. We modified this 
velocity field until a satisfactory match between observed and computed 
profiles was found. The advantage of this approach is that the velocities are 
not ``measured'' from the profiles, but modelled self-consistently along with 
the rest of the atmospheric parameters, and the region corresponding to each 
value of the velocity field is well determined, unlike with other methods such 
as, for example, the bisector.

\begin{figure}      
\begin{center}     
\includegraphics[width=9cm]{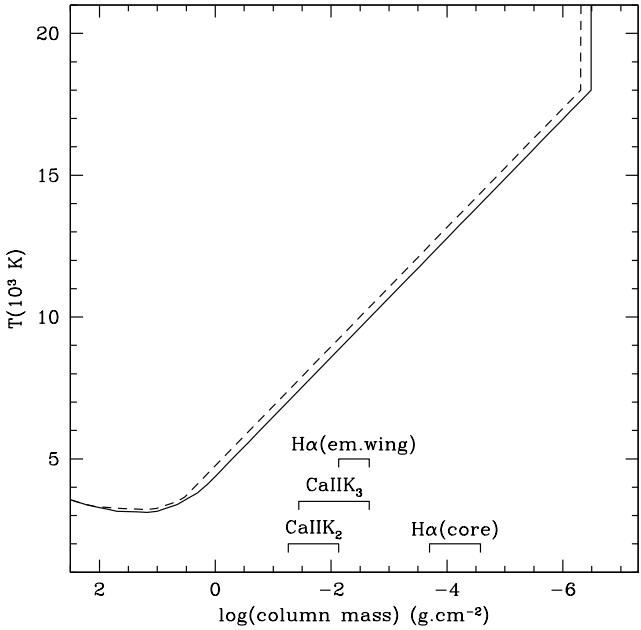}       
\caption{Our chromospheric models: model A (solid line), for
  the least active stars, and model B (dashed line) for the
  most active stars (37872 and 47606).These models do not include any
  velocity field. The approximate depth of formation of the 
  H$\alpha$ core and wings, and of the Ca {\sc ii} K$_2$ and
  K$_3$  line components are also indicated. 
}
\label{f:mods}      
\end{center}     
\end{figure}

\begin{figure}      
\begin{center}     
\includegraphics[width=9cm]{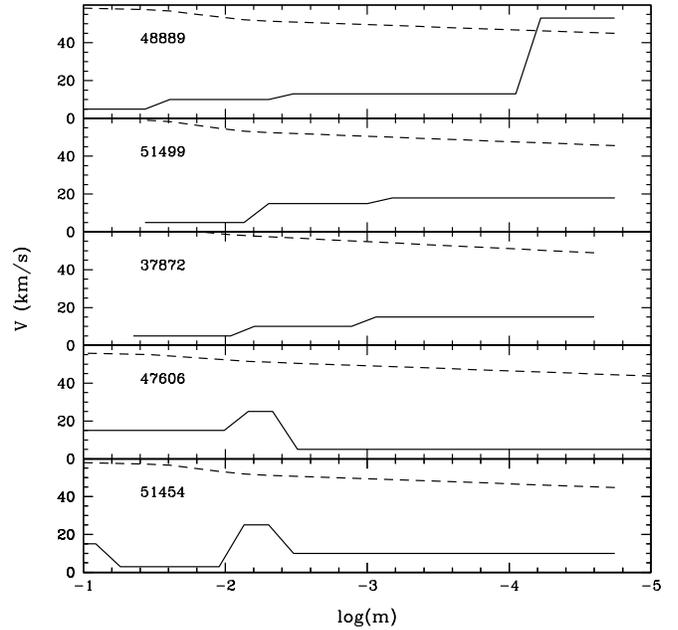}       
\caption{The velocity as a function of depth (solid line),
derived from the best match with the Ca K (innermost) and
H$\alpha$ (outermost) lines, compared to the corresponding
escape velocity (dashed line).         
}
\label{f:vel}      
\end{center}     
\end{figure}

Our two models are shown in Fig. \ref{f:mods}, together with the depth of 
formation of the lines we used as diagnostics, and the
velocities obtained are plotted in Fig. \ref{f:vel}.

\begin{figure}      
\begin{center}     
\includegraphics[width=9cm]{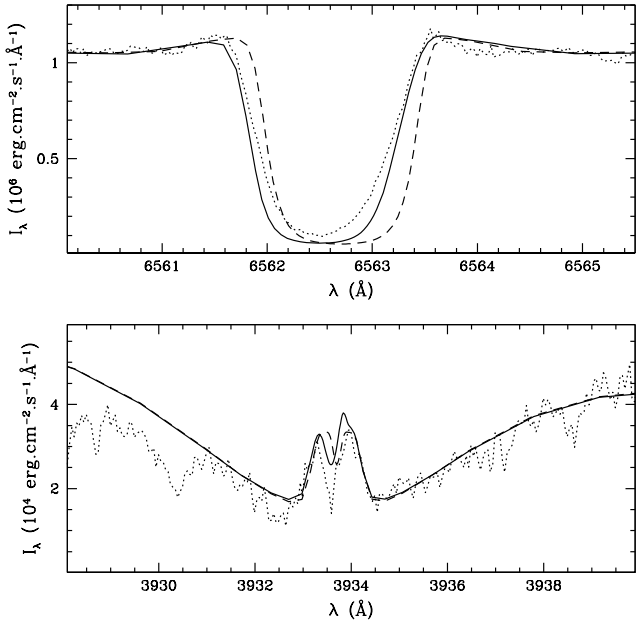}       
\caption{H$\alpha$ (top panel) and Ca {\sc ii} K (bottom panel) 
  observed profiles  (dotted lines) for the star 51499,
  compared with the computed profiles for our model A 
  with no velocity (dashed lines) and with  the velocity field shown in 
  Fig. 3 (solid lines).}
\label{f:prof9}      
\end{center}     

\end{figure}

\begin{figure}      
\begin{center}     
\includegraphics[width=9cm]{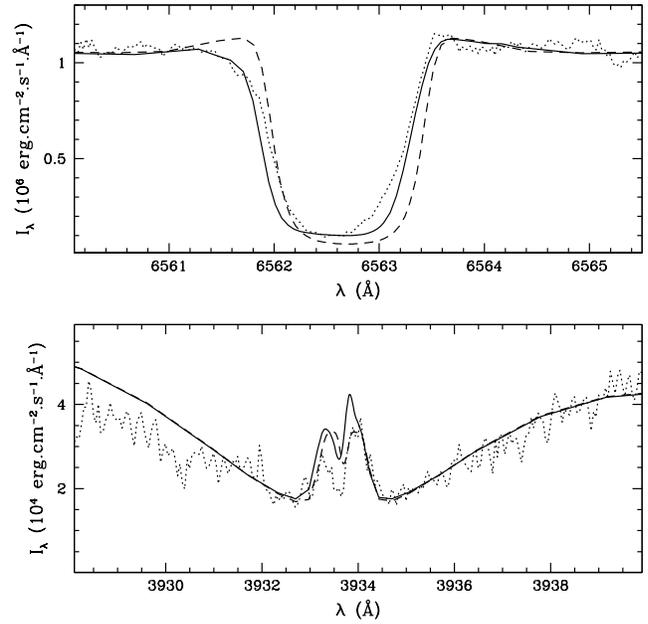}       
\caption{Same as in Fig. \ref{f:prof9} for the star 48889.
}
\label{f:prof8}      
\end{center}     
\end{figure}

\begin{figure}      
\begin{center}     
\includegraphics[width=9cm]{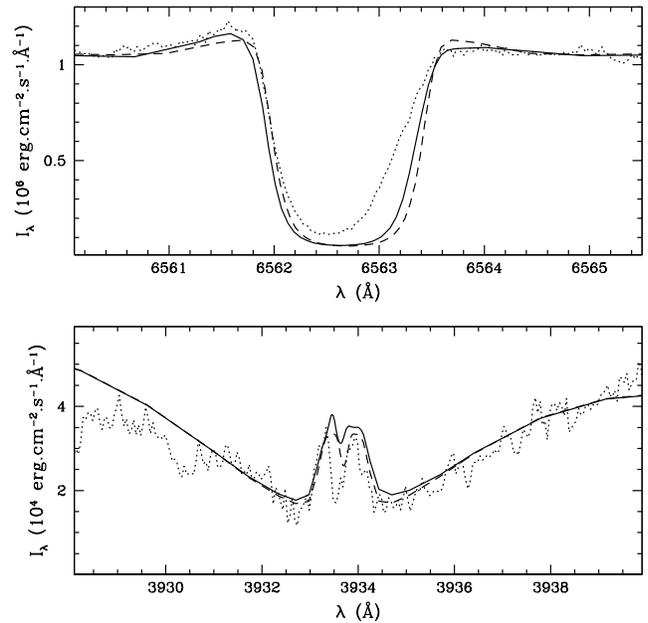}       
\caption{Same as in Fig. \ref{f:prof9} for the star 51454.
 }
\label{f:prof4}      
\end{center}     
\end{figure}

\begin{figure}      
\begin{center}     
\includegraphics[width=9cm]{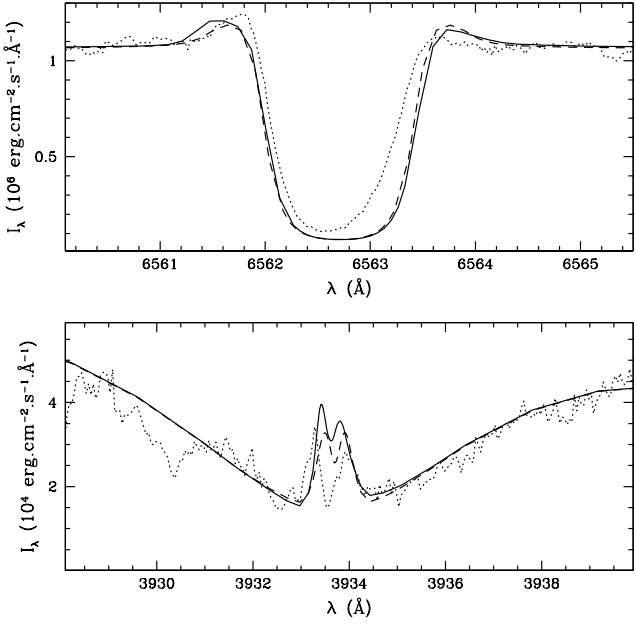}       
\caption{Same as in Fig. \ref{f:prof9} for the star 47606.
In this case we use model B.
}
\label{f:prof6}      
\end{center}     
\end{figure}

\begin{figure}      
\begin{center}     
\includegraphics[width=9cm]{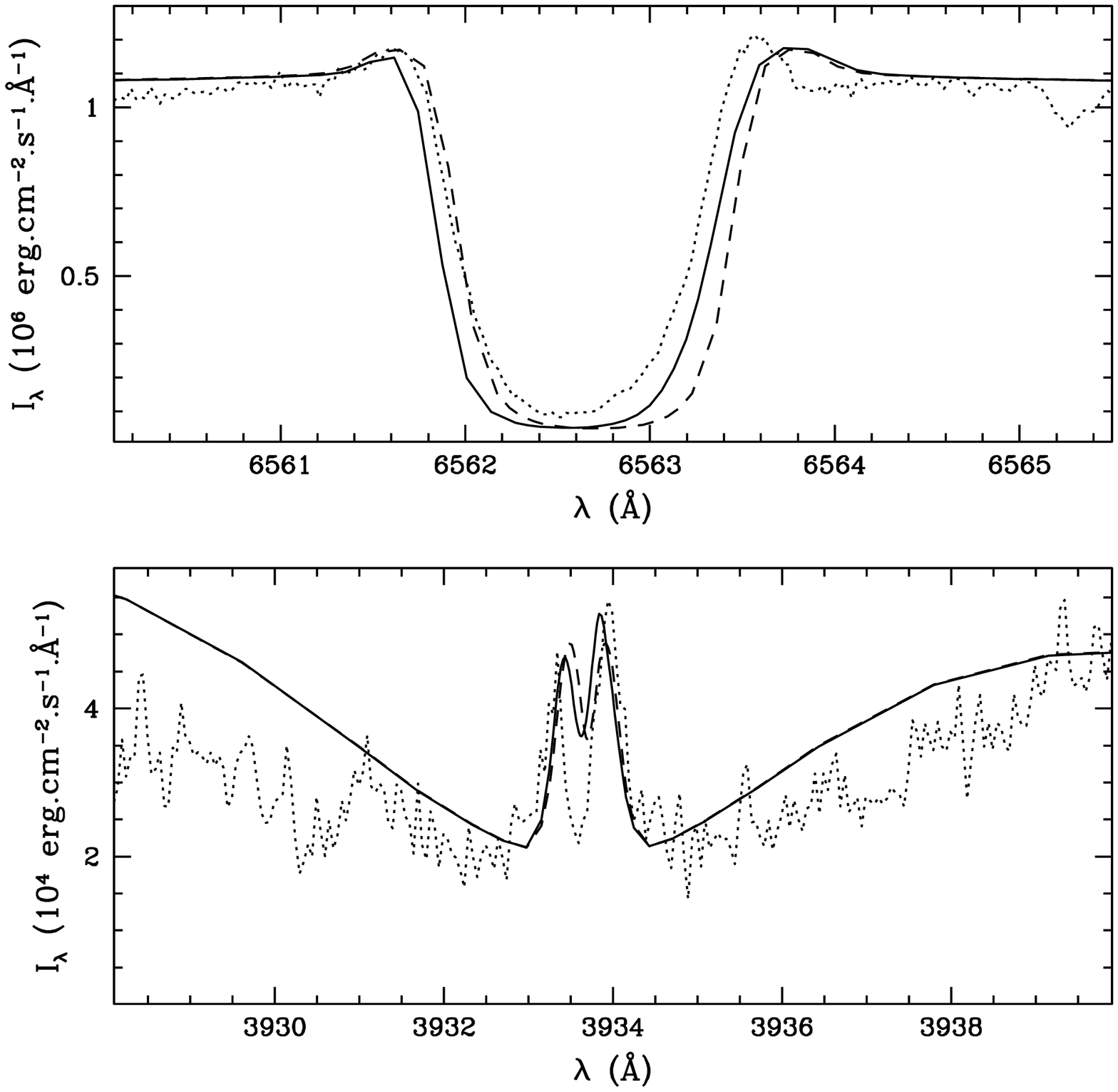}       
\caption{Same as in Fig. \ref{f:prof9} for the star 37872. 
In this case we use model B.
}  
\label{f:prof2}      
\end{center}     
\end{figure}

 The computed
H$\alpha$ and Ca {\sc ii} K line  profiles are compared with
the observations in Figs. \ref{f:prof9} to \ref{f:prof2}. 
The observed profiles have been reported to the continuum emission implied
by Kurucz's model and shifted by a quantity
corresponding to the individual stellar radial velocity 
(see Table 2 in C04), to report the profiles to their
wavelengths at rest. Also shown are the profiles computed
for the corresponding model without including the velocity
field, as a reference for clarity. These profiles have been
superimposed to those with velocity 
to highlight the asymmetries.

\begin{figure}  
\begin{center}     
\includegraphics[width=9cm]{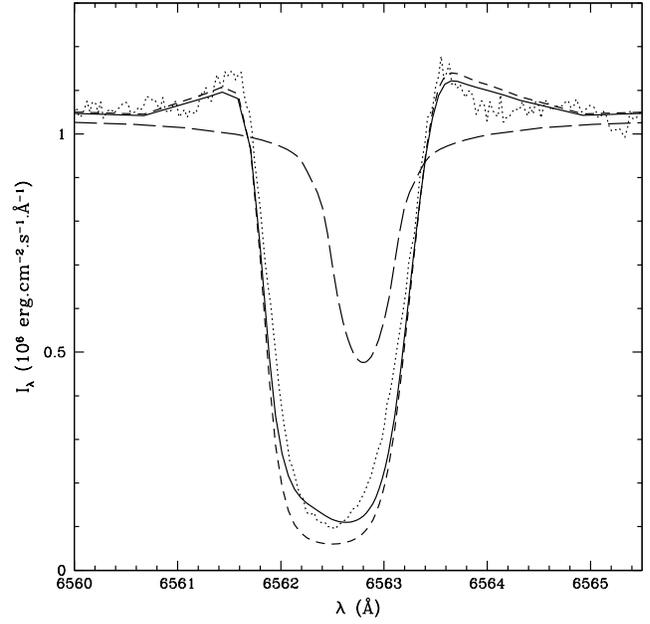}       
\caption{H$\alpha$ synthetic profiles computed for our model A 
  (short-dashed line), and for a model with no chromosphere (long-dashed lines). 
  The profile obtained combining 90\% of model A and 10\% of the 
  no-chromospheric model (solid line) gives a better fit to the observed 
  profile of star 51499 (dotted line) than the purely chromospheric profile.    
}
\label{f:fot}      
\end{center}     
\end{figure}

One can see in the figures that the agreement between computed 
and observed profiles is quite good. In fact, the main difference 
is that the H$\alpha$ computed profiles are broader and
slightly deeper in the line core.
We point out that in these calculations we are assuming an
homogeneous atmosphere, and that the broader and
deeper core shown by the computed profiles can be due to
inhomogeneities in the atmosphere. For example, in
Fig. \ref{f:fot} we show the H$\alpha$ profile 
computed from model A, and the profile obtained from an atmosphere 
with no chromospheric rise (and no velocity field),  
i.e. a purely photospheric profile.
Also shown is the profile obtained by combining 90\% of the profile computed
from model A and 10\% of the profile obtained from the model without 
chromospheric rise, as a way to simulate an inhomogeneus atmosphere. 
The resulting profile is narrower and more asymmetric in the center,
and fits much better the observations. We stress that the 
results discussed below are not affected by this disagreement, 
since they are mainly based on the asymmetries of the wings.   

Also, some of the Ca {\sc ii} photospheric profiles in Figures 5-8 
show a slight mismatch on the blue side of the line, and the model 
is slightly higher than the observed profile. 
This is most likely due to the presence of extra photospheric absorption 
lines in the spectra which are not included in the model,
like the Fe I lines at 3930.3 \AA ~and 3929 \AA. 
Some small residual tilt induced by the flat fielding process cannot be 
excluded (see e.g. the discussion in Pasquini 1992), but this is  not 
relevant to the following discussion. 
In modelling the Ca {\sc ii} K line we have been careful in matching the 
chromospheric core  intensity, asymmetry and width 
(Wilson-Bappu width, Wilson \& Bappu 1957). 
The models tend to produce a higher K$_3$ central intensity with 
respect to the observed profiles. This is also possibly due to the assumption
of an homogeneous chromosphere, since any contribution of a pure photospheric
line should be noticed mainly in the line center.   
We do not expect a significant contribution of interstellar Ca, since 
we do not see evidence of it in our spectra, and it could contribute 
to the Ca {\sc ii} core profile only if the velocity of the interstellar
medium were similar to that of the cluster ($\sim$ 100 km~s$^{-1}$).

In our models, the chromosphere starts ({\it i.e.} our model
departs from Kurucz' purely photospheric one) at log(m)=0.8
for the least active stars, and at 0.95 for the most active
ones. It is this fact that results in stronger emission wings
for model B. 
In both models, the slope of the chromosphere ({\it i.e.} $T$ vs
log(m),  with T in K and m in g.cm$^{-2}$) is --21000, 
and the ``transition region'' starts where
the temperature reaches 18000 K. 

Regarding the $T_{\rm min}$ region, we found the best 
agreement in the Ca {\sc ii} K${_1}$ minima with a slightly 
rounder temperature distribution, which does not show the sharp 
transition from Kurucz' model to ours that can be seen in 
Fig. \ref{f:exa}. 
In fact, the temperature for model B is slightly larger in
this region, to account for the more intense K$_1$ minima
seen in stars 37872 and 47606.

Regarding the slope of the chromospheric temperature vs mass, 
we found that this is the fundamental
parameter affecting the H$\alpha$ emission wings: a
shallower slope results in no emission wings, and
a steeper slope produces too large an emission. In other words, 
the emission wings are a direct sign of how steep the
chromospheric temperature rise is.

With respect to the upper end of the chromosphere, we point
out that the temperature distribution above $T \approx 10^4$
K does not affect our observed profiles. Therefore, the
influence of this parameter is related to the mass load of
the chromosphere, and not to the real temperature.  

\subsection{Velocity Fields and Mass Loss}

We see in the figures that the {\em presence}
of  emission in the H$\alpha$ wings can be reproduced
without considering any velocities. 
However, the asymmetries of these wings, and of other
features of the profiles, can only be reproduced with the
inclusion of velocity fields. In all cases, the
profiles including some velocity provide a better fit to the
data than those with zero velocity.  

The H$\alpha$ line core is the feature that forms higher up
in the atmosphere, and therefore we can only infer the
velocity up to the height where it is formed, about 
1 stellar radius above the photosphere. These are the
velocities $V_{exp}$ listed in Table \ref{t:list}. In  
Fig. \ref{f:vel} we show the full velocity field. For comparison, we 
also show the corresponding escape velocity, at each height 
Z from the center of the star, computed as V$_{esc}$=620\ (M/Z)$^{1/2}$,
where Z and the stellar mass M are in solar units. We
assumed a typical mass M=0.85 M$_{\odot}$ for all the stars. 

It can be seen that for one of the stars (star 48889) the 
expansion velocity at the most external point  
estimated from the H$\alpha$ profile is 
larger than the escape velocity at about 2 stellar radii
(from the center of the star), where the H$\alpha$ core forms. 
This velocity is reflected in the blue wing of the H$\alpha$
profile, which has an absorption feature whereas the red wing
has an emission peak. This is due to the displacement of the
line core, which is shallower in the profile with velocity.

This implies that, at least for this star, the material
above this level is escaping from the atmosphere. 
For the other stars, the expansion velocity we estimate for
the outermost point for which we have an indication, {\it
i.e.} where the H$\alpha$ core is formed, is smaller than the  
escape velocity at that depth. However, there is a mass outflow 
at this height, and the velocity is increasing in all cases,
while, of course, the escape velocity is decreasing. Furthermore,
there are no signs of a reversal in the velocity field,
{\it i.e.} there is no material falling back in. Therefore,
the most probable situation is that the velocity continues to
increase, until eventually it reaches the escape
velocity. The particularity of star 48889 is only that it
reaches this high level of velocity deep enough to be seen
in the H$\alpha$ profile.

We can estimate the mass loss rate for these stars using a 
simple formulation based on mass outflow:

$$\dot{M} = 1.3 \times m_H \times N_H \times 4\pi
  \times R^2 \times V\ $$

\noindent where $m_H$ and $N_H$ are the mass and the density of
hydrogen atoms, R is the distance from the stellar center
and V is the velocity of the outermost layer for which we
have a determination. The factor 1.3 takes care of the helium
abundance Y=0.23. We find that $N_H$ is between 
$\sim 1.3\times10^7$ and  $\sim 2\times10^7$ for all our 
stars (except 47606 for which the density is $\sim
1\times10^6$), at about R=2 R$_{\ast}$; 
the corresponding velocities range from
$\sim$ 10 to 50 km~s$^{-1}$ (see Table \ref{t:list} column 11). 
The values of $\dot{M}$ thus derived (and listed in column
12 of that Table) are a few times 10$^{-9}$ M$_{\odot}$
yr$^{-1}$ (ten times smaller for the star 47606). 

The rates expected from stellar evolution theory can be 
estimated using Renzini (1977) parameterisation (his eq. 6.10)

$$\dot{M} = 3\times10^{-11}\eta_{FPR}(\alpha/1.5)^{3.52}
  (10^3Z)^{-0.04}M^{-1.4}L^{1.92}\ $$

\noindent assuming $\eta_{FPR} = 6\times10^{-4}$ and $\alpha$=1 
(Renzini 1977), and mass M=0.85 M$_{\odot}$ and metallicity 
Z=1.23$\times10^{-3}$  (corresponding to [Fe/H]=--1.14). 
With these assumptions and 
approximations, the mass loss rate depends only on luminosity. 
Compared to the values listed in Table \ref{t:list} column 12, 
these ``evolutionary'' estimates are 2-10 times larger ($\sim$ 70 
times larger for star 47606).

For completeness, we have estimated the mass loss rates
according to eq. (4) of Schroeder \& Cuntz (2005)

$$\dot{M} = 1.98 \times 10^{-26} \times L_{\ast}R_{\ast}/M_{\ast} 
 \times T_{eff}^{3.5} \times (1+6.372/g_{\ast})\ $$

\noindent where $L_{\ast}$, $R_{\ast}$ and $M_{\ast}$ are stellar luminosity, 
radius and mass in solar units, and $g_{\ast}$ is the stellar surface gravity. 
This relation is an updated version of Reimers (1975)  
law $\dot{M}$ = $4\times10^{-13} \eta L/gR$ derived for
Population I K and M giants and supergiants, and includes an additional 
term related to the stellar temperature that seems to extend its validity 
also to metal-poor (Population II) stars. 
The values thus obtained range from 1.5 to 4.1
$\times$  10$^{-8} M_{\odot}$ yr$^{-1}$, and represent the 
{\em average} mass loss rates under the assumptions specified by Schroeder 
\& Cuntz for the mass loss mechanism (i.e. cool wind possibly related to 
Alfv\'en waves in the chromosphere or underneath).  
They are about ten times larger than the values we estimate
in the present study. These various estimates of mass loss rate may indicate 
the uncertainty of these determinations.

\subsection{Energy Requirements} 

To estimate the energy requirements to sustain the
chromosphere, {\it i.e.} to constrain the possible heating
mechanisms at work, we
compute the radiative cooling rate $\Phi$ (erg cm$^{-3}$ s$^{-1}$),
namely the net amount of energy radiated at a given depth by the
atmosphere, which is given by

$$ \Phi = 4 \pi \int \kappa_{\nu}\ (S_\nu-J_\nu)\ d\nu     $$
                      
\begin{figure}  
\begin{center}     
\includegraphics[width=9cm]{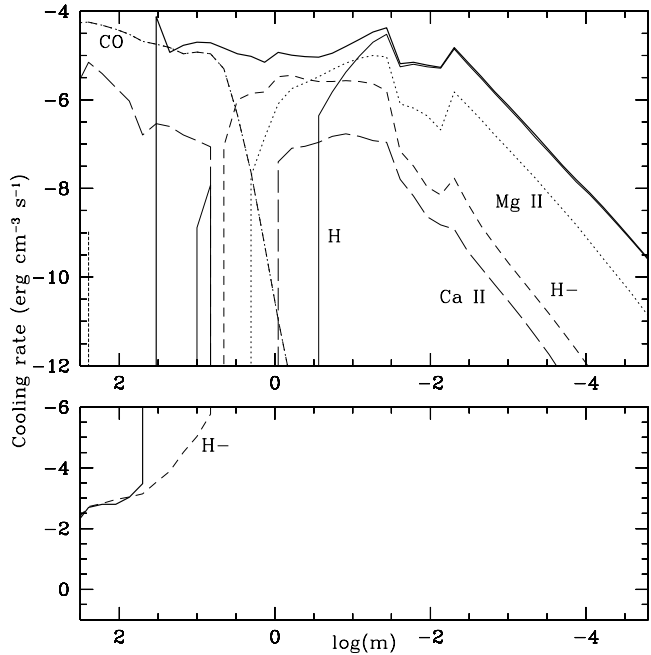}       
\caption{Log$ \Phi$, {\it i.e.} the total net radiative cooling rate, as a 
function of depth for our Model A (thick line), and the most important 
contributions to it.}

\label{f:cool}      
\end{center}     
\end{figure}

In this study we compute the contributions due to H$^{-}$,
H, He {\sc i}, Mg {\sc i} and {\sc ii}, 
Ca {\sc i} and {\sc ii}, Fe {\sc i}, Si {\sc i}, Na {\sc i},
Al {\sc i} and CO. The overall results and the
most important individual contributions for model A are presented in Fig.
\ref{f:cool}. A positive value implies a net loss of energy 
(cooling), and a negative value represents a net energy absorption 
(heating).

These rates can be compared with those for dwarf stars of
similar T$_{eff}$ (Mauas et al. 1997). It can be seen that
the energy required is, as expected, much lower for giant
stars: in the
mid-chromosphere, where it is largest, it reaches up to
10$^{-4}$ erg cm$^{-3}$ s$^{-1}$ for these giant stars, 5
orders of magnitude smaller than for the dwarfs.
However, since the particle density is also 5 orders
of magnitude smaller in giants, the energy {\it per
particle} is of the same order.

In the temperature minimum region the cooling rate is
negative. In this case there should be a missing cooling 
agent, a fact that was already noted for  dwarf stars. 
However, it should be pointed out that almost all cooling
is due to CO, which is the only molecule included in the
calculations. It seems reasonable to expect that, should
other molecules be computed in detail, the results would
show a similar behaviour as the CO calculations, bringing
this region closer to radiative balance. 

In the mid and high-chromosphere the energy balance is 
determined essentially by the hydrogen cooling rate. In this
region, the Mg {\sc ii} rate is also important, and it is an order
of magnitude larger than the one for Ca {\sc ii}.
Although it is not trivial to compare chromospheric non radiative 
losses in different lines, such as Mg {\sc ii} and Ca {\sc ii}, 
observed with different instruments, Pasquini \& Brocato (1992) 
analyzed the metal-poor stars observed by Dupree et al. (1994)
and found that chromopheric losses in the Mg {\sc ii} lines are a 
factor 2-3 higher than in the Ca {\sc ii}. This is, at least 
qualitatively, in agreement with the present analysis.

\section{Discussion and Conclusions}

We can summarize our conclusions as follows: 

\begin{itemize}

\item
The existence of emission in the Ca {\sc ii} K and H$\alpha$ lines
is of chromospheric origin with no need of a velocity field.
However, a rather steep rise of temperature with log(m) is needed 
to explain this emission, which is remarkably similar for all the stars
we studied.

\item
The asymmetries in the above line profiles, in particular 
for H$\alpha$, indicate the presence of velocity fields.   
These ``expansion'' velocities range from about 10 to  
50 km~s$^{-1}$, and provide clear evidence that some mass 
outflow occurs in these stars.  

\item
The rate of mass loss is estimated as a few 
10$^{-9}$ M$_{\odot}$ yr$^{-1}$, in good agreement with
previous estimates from the Mg {\sc ii} $k$ line (Dupree 
et al 1990b, 1994) and with the requirements of the stellar
evolution theory.  

\item
The energy per unit volume required to sustain the chromosphere 
is much smaller than the energy needed for a dwarf star of similar 
T$_{eff}$. However, the energy {\it per particle} is of similar order. 

\end{itemize}

\begin{acknowledgements} 

We would like to thank Dr. E.H. Avrett for interesting discussions and
comments. C.C. acknowledges the support of the MIUR (Ministero dell'Istruzione,
dell'Universit\`a e della Ricerca). PM acknowledges the Visiting Scientist 
programme at ESO-Garching.

\end{acknowledgements}

\end{document}